\documentstyle[12pt,axodraw,graphics]{article}  
\begin{document}
\pagestyle{empty}
\centerline{ \hfill          SLAC-PUB-8728
\footnote{Work supported by the Department of Energy, Contract 
DE-AC03-76SF00515}
}
\centerline{ \hfill          November 2000} 
%
\vskip 1cm  
%
\begin{center}
{\bf \large
The effects of spin fluctuations on the tunneling spectroscopy in High-$T_c$
Superconductors 
}
\vskip 1cm   
{C. L. Wu$^1$, Chung-Yu Mou$^{1,2}$, Darwin Chang$^{1,3}$}
\vskip 1cm
{\em $^1$Department of Physics, National Tsing Hua University, Hsinchu 30043,
Taiwan}\\
{\em $^2$National Center for Theoretical Sciences, P.O.Box 2-131, Hsinchu, Taiwan}\\
{\em $^3$Stanford Linear Accelerator Center, Stanford University, Stanford, 
     CA 94309, USA}\\

\end{center}
\date{\today}
\vskip 1.5cm  

\begin{abstract}
We investigate the effects of spin fluctuations on the
tunneling spectra of the NS junction. In the high junction resistance limit,
the dip/hump structure observed in ARPES data for the high Tc
superconductors is reproduced in a RPA treatment of the t-t'-J model. 
It is shown that the dip/hump structure weakens as doping increases
as reflected in the data.
In the other limit, we predict that the zero bias
Andreev peak can coexist with the dip/hump structure.
Furthermore, the c-axis tunneling spectra is found to 
be very similar to a recent STM data
once these fluctuations are included.
\end{abstract}

\{\small PACS numbers: 74.20.-z, 74.50.+r, 74.80.FP, 74.20.Mn \}



While a great details about the spectral function for
high-$T_c$ superconductors (HTS) have been
revealed by the angle resolved photoemission spectroscopy (ARPES)
\cite{arpes}, a complete description of the superconducting state requires
knowledge of the anomalous Green's function. Conventionally the tunneling
spectroscopy has been considered as one of the tools which can probe
the anomalous Green's function. In particular, measuring the
subgap conductance of a junction consisting of a normal metal and a
superconductor (NS) is the most convenient configuration for such a purpose.
To calculate the conductance in the NS configuration, Blonder, Tinkham, and
Klapwijk\cite{BTK}(BTK) developed a formalism 
using Bogoliubov-de Gennes (BdG) mean-field equations. The BTK theory has been
phenomenologically extended to investigate the tunneling phenomena in
various NS junctions\cite{BdG}. On the experimental side, even
though the d-wave BCS mean field theory captures some features of the superconducting
state, the recent high resolution data from both ARPES and STM gives a more delicate
picture. A new feature is the appearance of so called
peak/dip/hump structure, which is most clearly seen along the [100]
direction in the superconducting state for ARPES \cite{des,ran} and for STM \cite{DeWilde}.
It has been suggested\cite{shen} that it stems from the coupling of electrons to
the $\pi$ resonance observed in neutron
scattering studies\cite{neutron}. This
idea has been further explored both qualitatively\cite{aba} and
quantitatively\cite{Li00}, confirming its validity. In addition to the
peak/dip/hump structure, there are also indications that the quasi-particle
peak seems not to be resolution limited\cite{Ding00}. These features
indicate the need for a tunneling theory that includes the
effect of fluctuations.

In this work, we investigate the effects of the spin
fluctuations on the tunneling conductance spectra along the [100] and
[001] (c-axis) directions\cite{note0}. 
By using the Keldysh formulation, we first
demonstrate that in the high junction resistance limit, 
the dominant contribution comes from
the spectral function and the peak/dip/hump structure exists
in this limit.
This structure results from collective spin fluctuations and weakens as the doping $\delta$ increases as reflected 
in some of the recent data, such as in Ref.\cite{miyakawa}. 
Since spin excitations are gapped, they induce little qualitative 
change in the subgap region. Instead, their main effect is to 
redistribute the spectral weight and thus changes the relative
strength among currents due to difference tunneling processes. 
Therefore, to investigate the other limit when
the subgap is dominated by the Andreev reflection
, we define the {\it optimum matching}
as the condition when the
zero bias Andreev conductance peak reaches maximum. 
This corresponds to the $Z=0$\ case in the BTK theory. 
Under this condition, we show that 
the Andreev peak can coexist with the dip/hump structure,
which should be observable in the recent future.
Base on
our analysis, we will also give a possible explanation on a recently observed
c-axis STM data \cite{pan}, which shows an unexpected step like feature in the
negative bias in addition to the peak/dip/hump structure.

We start by considering a junction consisting of a 2D normal metal on the
left ($L$) hand side ($-\infty <x\leq 0$) and a 2D superconductor ($a\leq
x<\infty $, $a$ is the lattice constant) on the right ($R$) hand side,
governed by the Hamiltonian $H_L$ and $H_R$ respectively.
The tunneling Hamiltonian that connects the surface points 
at $x=0$ and $x=a$
is given by
$ H_T=\sum_y t(|y_L-y_R|)(c_L^{\dagger }c_R+c_R^{\dagger }c_L)$,
where the summation is over lattice points along the interface, chosen to be
in the y direction. We consider the simplest case when the lattice
points along the interfaces are equally spaced and match the bulk lattice of
the metal and the superconductor. The superconductor is assumed to have a
square lattice with one of the axis parallel to the x direction. 
The total grand Hamiltonian is then given by
$ K =H_L-\mu _LN_L+H_R-\mu _RN_R+H_T$,
where $\mu _L$ and $\mu _R$ are the chemical potentials and their 
difference $
\mu _L-\mu _R$ is fixed to be the voltage drop $eV$ across the junction.

The tunneling current can be calculated
perturbatively by using the Keldysh formalism\cite{Keldysh}.
This approach was previously applied successfully to study a 
number of tunneling problems\cite{Yeyati,Hu}. We shall follow
Ref.\cite{Yeyati} and neglect the vertex corrections
of $H_T$. The perturbation series in $H_T$ can
be then summed exactly. 
The contribution to the differential conductance 
$G = d I / dV$ can be classified
into four terms due to different tunneling processes\cite{Yeyati}: $G_1$
is due to particle to particle tunneling; $G_2$ is
due to particle to particle tunneling with pair
creation/annihilation as the intermediate state; $G_3$ is
due to particle to hole tunneling; and $G_A$ is the Andreev conductance.
Because $H_T$ is a tight binding model, all the Green's
functions have to be replaced by
the surface Green's functions that connect different
points on the junction. Thus there is an extra integration
over $k_y$ and associated with each surface Green's function, 
there is a $t(k_y)$ factor.
The function $t (k_y)$
characterizes the spread of the electron wave function along y direction
when hopping across the junction and can be generally expanded in a cos Fourier series.
Note that the bare surface Green's function, without
being renormalized by $H_T$, is a 2$\times $2
matrix $\hat{g}_0(\omega ,k_y)$ in Nambu's notation\cite{Yeyati},
and its relation
to the bare bulk Green's function $\hat{G}_0
(\omega ,k_x,k_y)$ is given by $\hat{g}_0(\omega
,k_y)=\frac 2\pi \int_0^\pi dk_x\sin ^2(k_x) \hat{G}_0(\omega,k_x,k_y)$, where the
momentum $k$ is in unit $1/a$.

The relations of conductance $G_{\alpha}$ to the Green's
functions can be best demonstrated in the limit when the metal is
approximated by its bandwidth $t_L$ with a constant density of
state, i.e., $g_L(\omega ,k_y)=-i/t_L\times {\rm I}$,
where ${\rm I}$ is a unit matrix. 
This avoids complications due to the band structure from the
metal side. In this case,
when $t (k_y)=t$ is a constant, the only dimensionless parameter
is $\lambda \equiv t^{2}/(t_Lt_R)$, where $t_R$
is the hopping scale of the superconducting side.
The junction conductance is then of order $\lambda\ e^2/\hbar$.
For small $\lambda$,
$G_1(V)$ is $O(\lambda )$ and is
simply proportional to the single
particle density of state\cite{DOS}. 
Similarly, $G_2(V)$ is of order $O(\lambda ^2)$ and probes $
\int dk_y\left[
\mathop{\rm Im}
(g_{0R,12}^r)\right] ^2$, $G_3(V)$ is of the 
order $O(\lambda ^3)$ and probes $\int
dk_y\rho _{R,22}\,|g_{0R,12}^r|^2$. 
Here 1 and 2 are indices for Nambu notations
, Green's functions with the index $r$ are retarded
and ${\rho}$ is the
spectral function.
Since $G_3$ is subdominant to $G_2(V)$
, $G_2(V)+G_3(V)$\ is negative. Finally, the Andreev conductance 
is $O(\lambda ^2)$ and probes 
$
\int dk_y\left[
(g_{0R,12}^r)\right] ^2$.
Overall speaking, for small $\lambda $, the
total conductance is dominated by $G_1$, corresponding to the large $Z$ limit
of the BTK theory. In the other limit when $G_A$ dominates in 
the sub-gap region,
the situation is more subtle. For s-wave BCS superconductors,
the analytic mapping from $Z$ to $\lambda$ was obtained in 
Ref~\cite{Yeyati}.
The optimum matching ($Z=0$) does not occur at large $\lambda$
because the mapping is not monotonic.
In general, such analytic mapping does not exist 
, we shall resort to numerics to find the optimum matching condition.

We first consider 
a simple d-wave BCS superconductor described by
\begin{equation}
H_R^m=\sum_{k\sigma }\epsilon _kc_{k\sigma }^{+}c_{k\sigma }-\sum_k\Delta
_k(c_{k\uparrow }^{+}c_{-k\downarrow }+h.c.),  \label{mean}
\end{equation}
where the dispersion $\epsilon _k=-2t_R[\cos (k_x)+\cos (k_y)]$, and
the gap $\Delta (k)=\Delta_R [\cos (k_x)-\cos (k_y)]$. 
The metal side has the same Hamiltonian with $\Delta_L=0$.
Under the optimum matching condition,
one obtains
a single peak (the Andreev peak) in the total conductance near
$V=0$\cite{Hu,note0}.
The Andreev peak obtained here is a result of
subtle balance between $G_1+G_A$ and $G_2+G_3$.
In fact, the effect of $G_2+G_3$ is to bring down 
the quasi-particle peaks in $G_1(V)$
so that a single peak is manifested. As we shall see,
such simple realization of the Andreev peak does not always
happen in real high-Tc systems due to spin fluctuations.

To include the spin fluctuations, we shall work 
with the 2D $
t-t^{\prime }-J$ model.
In the slave-boson method, 
the physical electron
operators $c_{i\sigma }$ are expressed by slave bosons $b_i$ carrying the
charge and fermions $f_{i\sigma }$ representing the spin; $c_{i\sigma
}=b_i^{+}f_{i\sigma }$. The mean-field $d$-wave SC state is characterized by
the order parameters $\Delta_0=\langle f_{i\uparrow }f_{j\downarrow
}-f_{i\downarrow }f_{j\downarrow }\rangle $, $\chi_0=\sum_\sigma
\left\langle f_{i\sigma }^{+}f_{j\sigma }\right\rangle $ and the condensate
of bosons $b_i\rightarrow \left\langle b_i\right\rangle =\sqrt{\delta }$.
Eq.(\ref{mean}) is then the Hamiltonian for the spinons, 
$f_i$, with dispersion 
$\epsilon _k=-2(\delta t_R+J^{\prime}\chi _0)[\cos (k_x)+\cos (k_y)]-4\delta
{t_R}^{\prime }\cos (k_x)\cos (k_y)-\mu_R $ and $\Delta_R =2J^{\prime
}\Delta_0$, where $J^{\prime }=3J/8$. We shall 
adopt
the following numerical values $t_R=2J$, 
${t_R}^{\prime }=-0.45t_R$, and $J=0.13$eV\cite{Li00,Tanamoto92}.
The mean-field parameters $\chi_0$, $\Delta_0$ and the chemical potential 
$\mu_R $ for different doping $\delta $ are obtained 
from a self-consistent calculation\cite{Li00}. 
Next we include the spin fluctuations by 
perturbing around the mean field Hamiltonian $H_R^m$, i.e., we write $
K_R=H_R^{m}+H^{\prime }$, and treat $H^{\prime }$ as a perturbation. 
In order to account for the $\pi$ resonance\cite{neutron} 
as well as many other effects of spin fluctuations, 
we calculate the spin susceptibility in a modified 
random-phase approximation (RPA) as defined in Ref.\cite{Li00,Tanamoto92}.  
The usual RPA sums over selected sets of graphs 
for the spin susceptibility $\chi$ as shown in 
Fig.1(a) and gives rise to $\chi ({\bf q},\omega )
=\chi _0({\bf q},\omega)/[1+\alpha J({\bf q})
\chi _0({\bf q},\omega )]$ with $\alpha =1$.  
Here, $J({\bf q})=J(\cos q_x+\cos q_y)$, 
$\chi _0({\bf q},\omega )$ is the {\it unperturbed} spin
susceptibility due to the spinon bubbles
and the $\pi$
resonance emerges as the pole of the denominator.  
In the current approach, 
$\alpha $ 
is not one and is considered as a phenomenological parameter
whose value is chosen such that 
the AF instability occurs right at
the experimental observed value $\delta =0.02$.
For the material parameters we adopt, $\alpha$ is $0.34$\cite{Li00}.

The inelastic 
scattering of electrons off the spin fluctuations is
taken into account by incorporating $\chi$ into the self energy
of the spinons in the lowest-order approximation.
In the SC state, there are two different
self-energies $\Sigma _{s}$ and $\Sigma _{w}$ as shown in Fig.1(b) and (c).
The Green's function for spinons is calculated by $G_{f}({\bf k}
,\omega )=[G_{f0}^{-1}({\bf k},\omega )+(\Delta _{k}+\Sigma
_{w})^{2}G_{f0}^{-1}(-{\bf k},-\omega )]^{-1}$ with $G_{f0}({\bf k},\omega
)=[i\omega -\epsilon _{k}-\Sigma _{s}({\bf k},\omega )]^{-1}$. 
Since bosons 
condense, the physical electron Green's function can be simply
obtained by $G({\bf k},\omega ) = \delta G_{f}({\bf q},\omega )$, i.e.,
only the dynamics of spins is considered. 
Following previous prescriptions, one then obtains
the surface Green's functions 
and thus the various conductance.
The truncation to the lowest order cannot really be 
justified rigorously so far.  
Its merit rests mainly upon its simplicity and its usefulness 
in previous applications to problems related to spin fluctuations\cite{Li00}.
These studies indicate that it
has captured the main features of ARPES data along [100] direction and
for other directions it also reproduces the observed $\cos (6\theta )$
deviation from the pure $d$-wave\cite{Li00}.
Here we shall examine its validity against tunneling data.
Note also that we had neglected the spatial dependence 
of the pair potential, which is generally 
considered not important in [100] and [001] 
directions.

We first analyze small $\lambda$ limit. 
Fig. 2 shows the total conductance with RPA correction for various dopings. 
The positions of the peak and hump are seen to scale weakly with doping. 
When doping increases, the height of peak increases with doping,
in consistent with experiments\cite{miyakawa}, at the same time, 
the width of the peak increases and tends into the hump region so 
that the hump is smeared out in slightly overdoped region.
Another feature which can also be observed in the data is that the
dip/hump feature at positive bias is always weaker.
The precise reason behind
them can be traced back to the underlying structure of $\epsilon_k$. In
fact, detailed analysis\cite{note0} shows that the band edge extends to higher
positive bias so that the dip/hump is smeared, while the band edge for
negative bias essentially stays at small bias, leaving the dip/hump
unsmeared.

We now numerically identify the optimum matching condition so that the zero
bias Andreev conductance peak can be manifested best. For each $k_y$,
we compute the optimal value $t_{opt}$ such that the Andreev
conductance at $V=0$ reaches maximum. The resulting
$t_{opt} (k_y)$ can be approximated by
\begin{equation}
t_{opt} (k_y)=a_0+a_1\cos (k_y)+a_2\cos (2k_y)  \label{fit}
\end{equation}
This implies that including next nearest neighbor hopping along the junction
is necessary. However, the forward hopping $a_0$ still dominates
(for instance, when $\delta=0.12$, we obtain $a_0=2.41$,
$a_1=-0.44$, and $a_2=0.34$). In Fig.3,
we show the optimal manifestation of the Andreev peak for different doping.
The metal side is modeled by a simple tight-binding model on the square lattice.
We see that the dip/hump structure coexists with the Andreev peak. Fig. 4 
shows a
similar plot but now the density of state of the metal side is a constant.
In this case, the Andreev peak never out wins the quasi-particle peaks 
resulted from $G_1(V)$ so that a plateau is
observed. In both cases, the trend of the dip/hump structure with doping is
consistent with what is found in Fig.2.  
All these qualitative features should be experimentally 
verified in the future as a test of the mechanism of the spin 
fluctuations.  

To further test this particular RPA approach, we compute the 
c-axis tunneling spectrum.
Fig. 5 shows our numerical results, in comparison to the recent STM
curve by Pan {\it et. al.} \cite{pan}. It is quite encouraging that two
curves are very similar in the shape. In particular, the step around 45
$mV$ is reproduced in the RPA approach at slightly larger bias. This
step results from the band edge, which, as we mentioned, essentially
stays at small bias as one changes doping.

To summarize, we have analyzed the effects of 
spin fluctuations on the SN junction using the Keldysh 
formulation and a modified random phase approximation. 
The peak/dip/hump structure is reproduced and 
we show that it disappears gradually as one 
goes to slightly overdoped region. 
Using the same formulation, we predict that 
the dip/hump structure can coexist with 
the zero bias Andreev peak in optimal matching conditions. 
Our analysis on the c-axis tunneling shows good qualitative 
agreement between this approach and the recently observed STM tunneling curve.

It is our pleasure to thank Profs. N. C. Yeh, C.R. Hu and T. K. Lee for
useful discussions. DC also wish to thank SLAC Theory Group for hospitality.  
This research was supported by NSC of Taiwan.


\newpage

\vspace{1cm}

\section{FIGURE CAPTIONS}

Fig.1 Feynman diagrams for (a) the spin susceptibility and (b)
and (c) the lowest order contributions to the self-energy from
spin fluctuations.

Fig.2
The total conductance with RPA correction in the
tunneling limit. Here the metal is modelled by a constant density of state
with $\lambda = 0.05$.

Fig.3
The optimal manifestation of the Andreev peak with a
square lattice ($t_L=1.0$) for the metal side.

Fig.4
The optimal manifestation of the Andreev peak with
constant density of state ($t_L = 1.0$) for the metal side.

Fig.5
The density of state for c-axis tunneling for $
\delta =0.12$.
Inset: The STM tunneling curve observed by Pan et al\cite{pan}.

\newpage

\begin{figure*}[tbp]
\centerline{\epsfxsize=6.0in \epsfysize=8.0 in
\epsffile{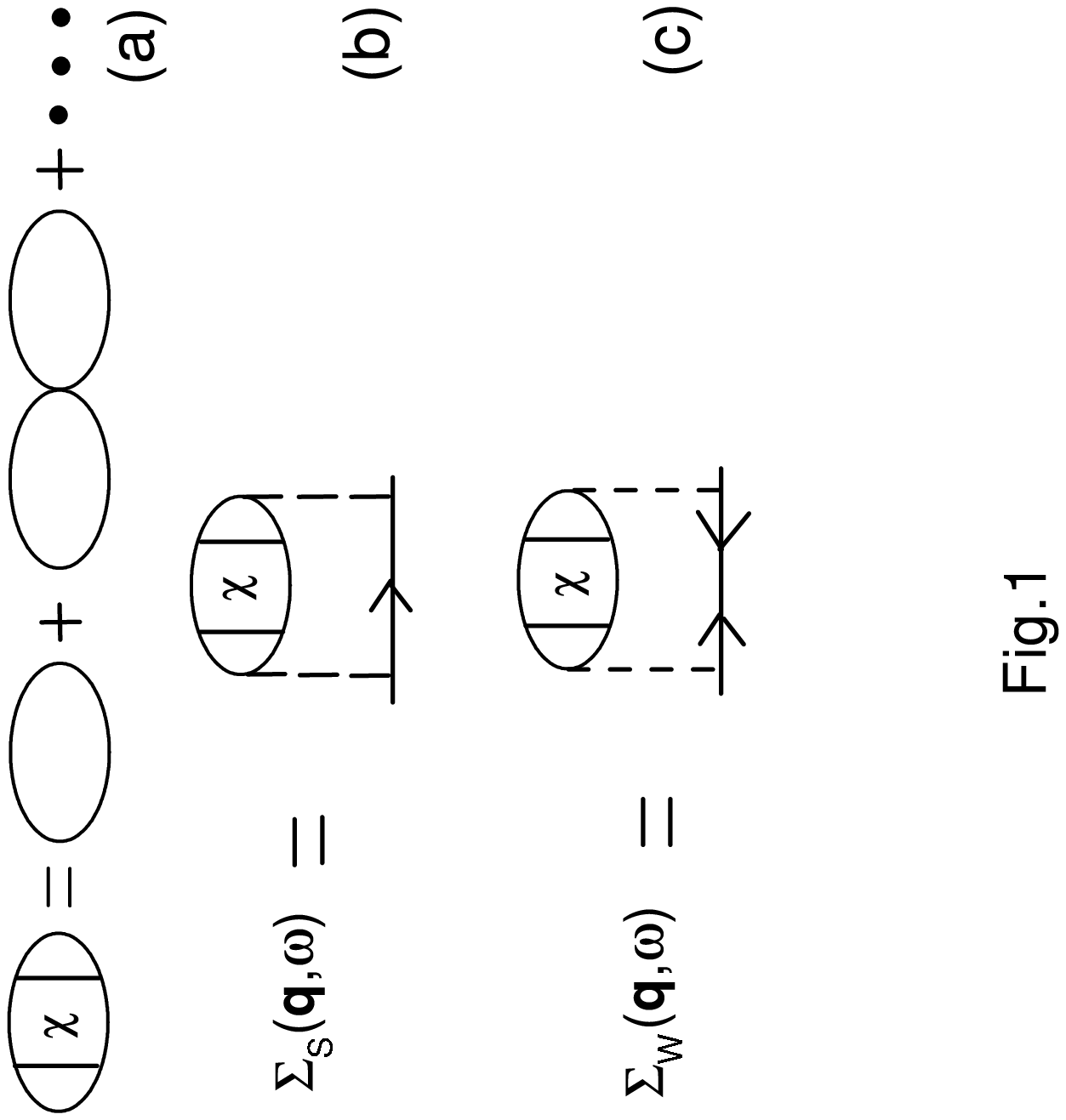}}
\end{figure*}

\newpage

\begin{figure}[tbp]
\centerline{\epsfxsize=6.0in \epsfysize=8.0 in
\epsffile{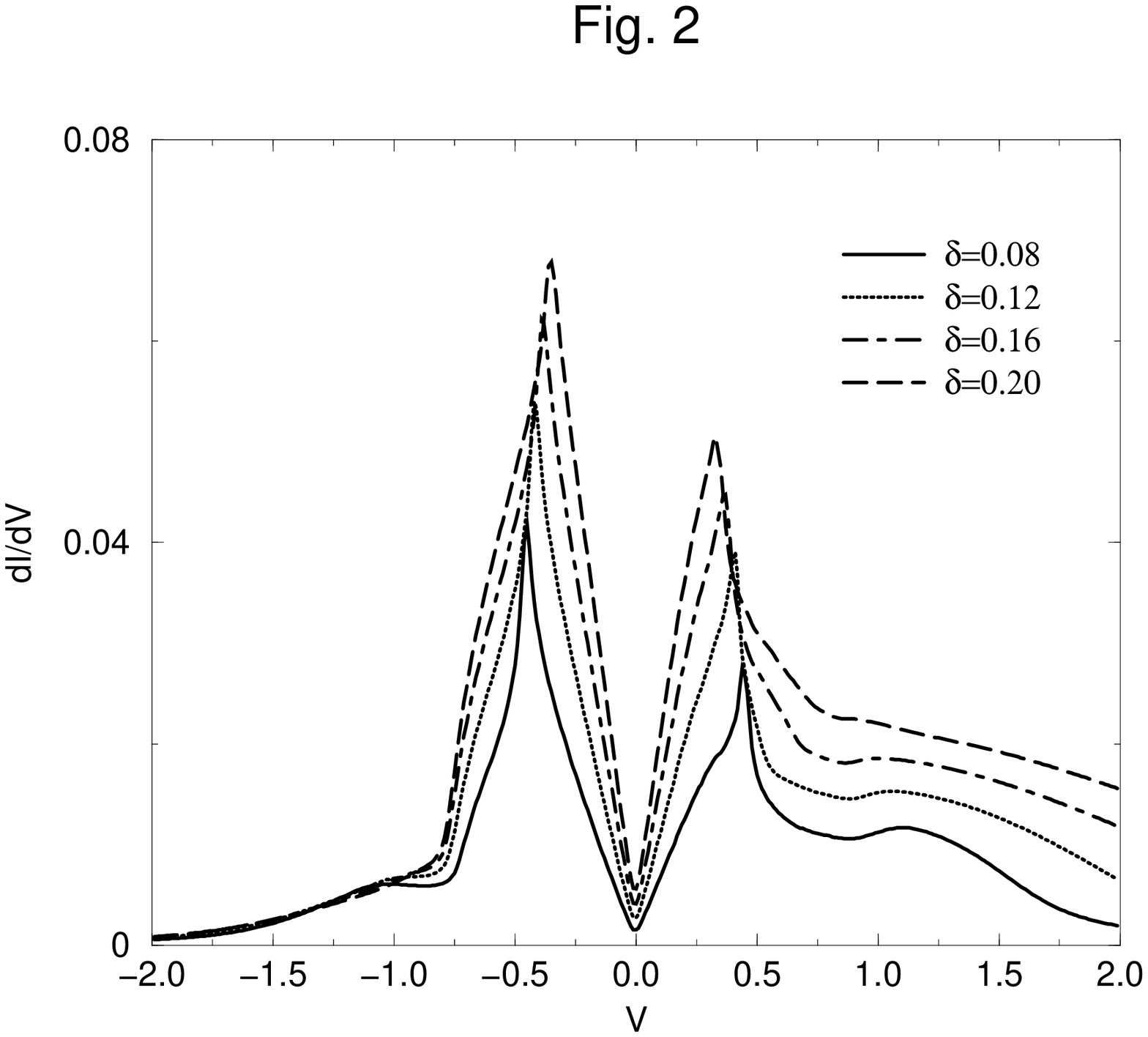}}
\end{figure}

\newpage

\begin{figure}[tbp]
\centerline{\epsfxsize=6.0in \epsfysize=8.0 in
\epsffile{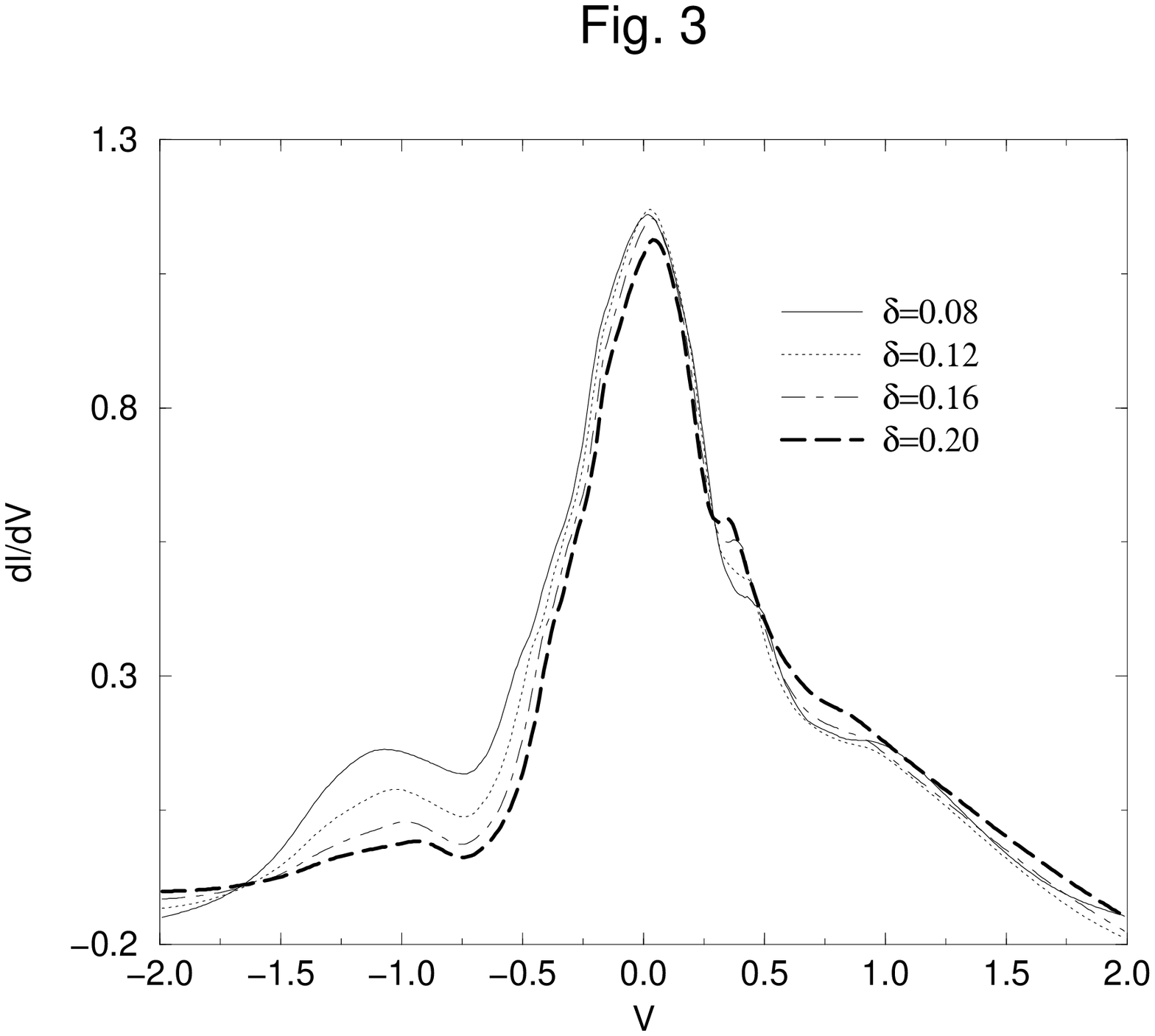}}
\end{figure}

\newpage

\begin{figure}[tbp]
\centerline{\epsfxsize=6.0in \epsfysize=8.0 in
\epsffile{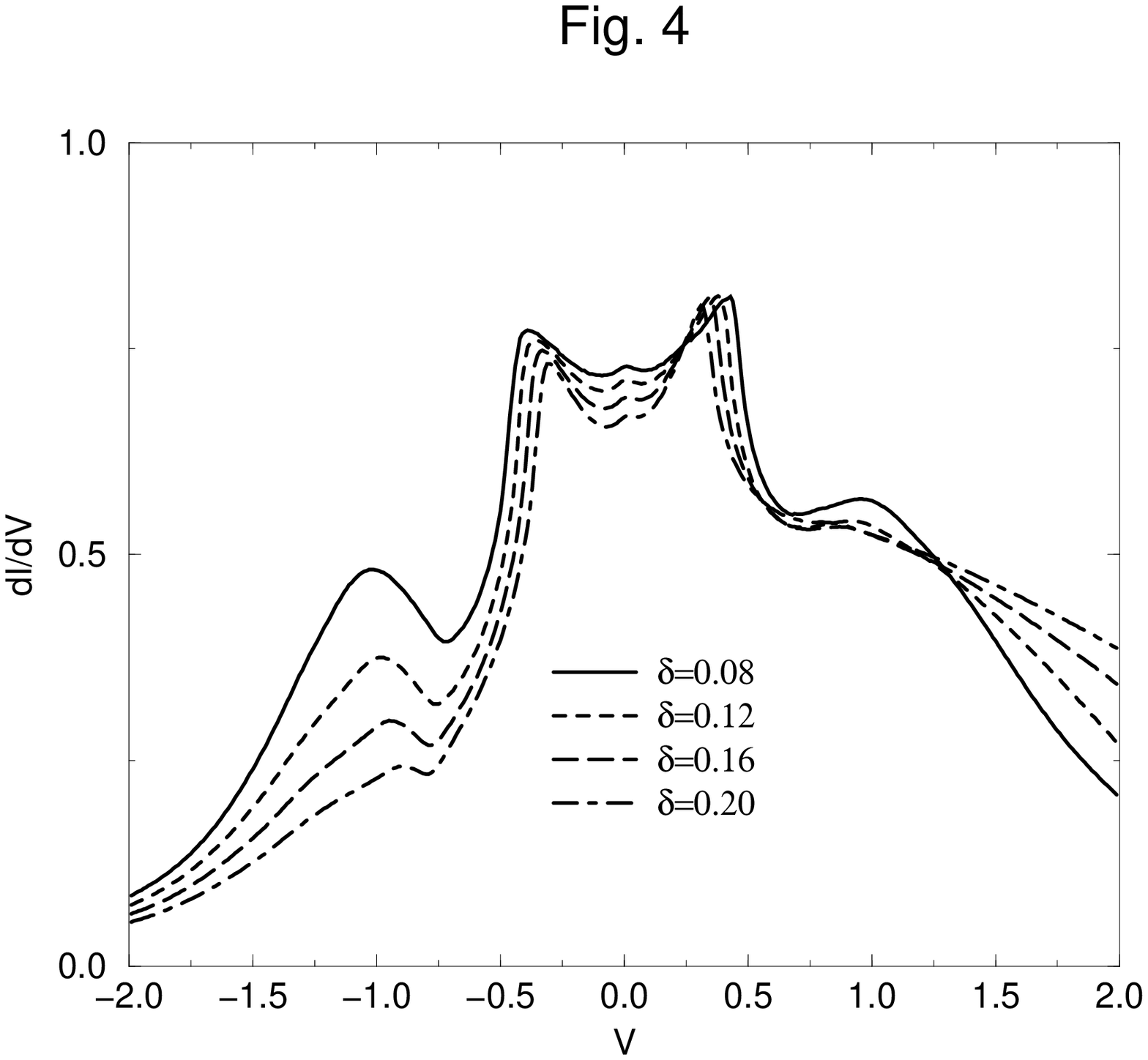}}
\end{figure}

\newpage

\begin{figure}[tbp]
\centerline{\epsfxsize=6.0in \epsfysize=8.0 in
\epsffile{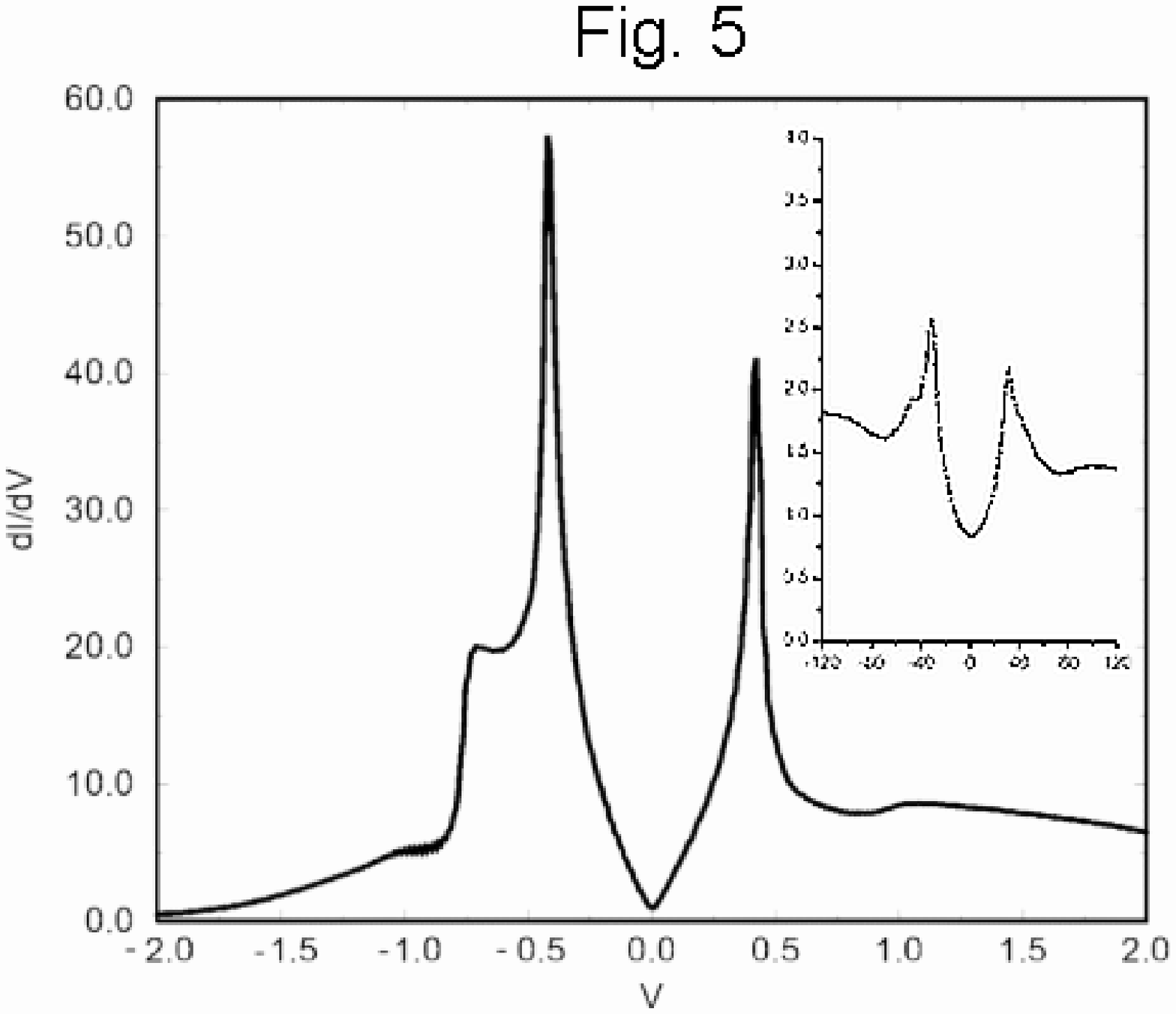}}
\end{figure}


\begin{thebibliography} {99}  
\bibitem{arpes}  See
recent review M. Randeria and J.C.Campuzano cond-mat 9709107.
%
\bibitem{BTK}  G. E. Blonder, M. Tinkham, and T. M. Klapwijk, Phys. Rev. B
{\bf 25}, 4515 (1982).

\bibitem{BdG}  See for example, Y. Tanaka and S. Kashiwaya, Phys. Rev. Lett.
{\bf 74}, 3451 (1995).

\bibitem{des}  D. S. Dessau, {\it et al.}, Phys. Rev. Lett. {\bf 66}, 2160
(1991).

\bibitem{ran}  M. Randeria, {\it et al.}, Phys. Rev. Lett. {\bf 74}, 4951
(1995).

\bibitem{DeWilde}  See for example, Y. DeWilde, {\it et al.,} Phys. Rev.
Lett. {\bf 80}, 153 (1998).

\bibitem{shen}  Z. -X. Shen and J. R. Schrieffer, Phys. Rev. Lett. {\bf 78},
1771 (1997); M. R. Norman and H. Ding, Phys.Rev.B {\bf 57}, R11089, 1998.

\bibitem{neutron} See for example, J. Rossat-Mignod {\it et al.}, Physica C {\bf
 185-189}, 86
(1991).

\bibitem{aba}  A. Abanov and A. V. Chubukov, Phys. Rev. Lett. {\bf 83}, 1652
(1999).

\bibitem{Li00} J. Brinckman and P. A. Lee, Phys. Rev. Lett.
{\bf82},2915, (1999); J. X. Li, C. -Y. Mou, and T. K. Lee,
Phys. Rev. B{\bf \ 62},
640 (2000).

\bibitem{Ding00}  H. Ding et al. cond-mat 0006143.

\bibitem{note0} It will be shown in a separate publication 
that when appropriately modifying the the relation of 
the surface Green's functions to the bulk Green's functions, 
the current formalism is also applicable to the 
[110] direction, where the ZBCP is observed (see Ref.~\cite{BdG};
C. R. Hu, Phys. Rev. Lett. {\bf 72}, 1526 (1994); S. Kashiwaya, Rep. Prog.
Phys. {\bf 63}, 1641 (2000) and references therein.)

\bibitem{miyakawa}  N. Miyakawa et al. Phys. Rev. Lett. {\bf 83}, 1018 (1999).

\bibitem{pan}  S. H. Pan et al. cond-mat/0005484.

\bibitem{Keldysh}  L. V. Keldysh, Sov. Phys. JETP {\bf 20}, 1018 (1965).

\bibitem{Yeyati}  
J. C. Cuevas, A.
Martin-Rodero, and A. L. Yeyati, Phys. Rev. B {\bf 54}, 7366 (1996).

\bibitem{Hu}  X. Z. Yan, H. Zhao, and C. R. Hu, cond-mat/0001170.

\bibitem{DOS}  For tunneling along [100] direction, the density of state
has to include an extra $\sin ^2(k_xa)$ factor due to the boundary condition
at $x=a$. This is different from the c-axis STM measurement, where the usual
density of state without $\sin ^2(k_xa)$ factor is measured.

\bibitem{Tanamoto92} Early RPA work can be found 
in H. Tanamoto et al. J. Phys. Soc. Jpn. {\bf 61}, 1886, (1992); 
{\it ibid} {\bf 62}, 717, (1993); {\it ibid} {\bf 62}, 1455, (1993).
Our parameters are different from those used in these references, however, 
they yield similar Fermi surfaces in the mean field level 
and therefore similar qualitative consequences.
\end{thebibliography}
\end{document}